\documentclass[notoc,11pt,paper]{JHEP3}
\date{March 2008}
\usepackage{epsfig}
\usepackage{latexsym}
\usepackage{amssymb}
\usepackage{booktabs}
\newcommand{\be}{\begin{equation}}
\newcommand{\ee}{\end{equation}}
\newcommand{\ba}{\begin{eqnarray}}
\newcommand{\ea}{\end{eqnarray}}
\newcommand{\bi}{\begin{itemize}}
\newcommand{\ei}{\end{itemize}}

\newcommand{\half}{{\textstyle\frac{1}{2}}}

\newcommand{\quarter}{{\textstyle\frac{1}{4}}}

\newcommand{\<}{\langle}
\renewcommand{\>}{\rangle}
\newcommand{\eq}{Eq.~}

\newcommand{\fig}{Fig.~}

\newcommand{\la}{\label}

\newcommand{\txts}{\textstyle}

\newcommand{\Nt}{N_{\tau}}

\title{Glueball matrix elements: a lattice calculation and applications}

\author{
Harvey~B.~Meyer\\
Center for Theoretical Physics\\
Massachusetts Institute of Technology\\
Cambridge, MA 02139, U.S.A.\\
E-mail: \email{meyerh@mit.edu}
}

\preprint{MIT-CTP 3973}

\abstract{
We compute the matrix elements of the energy-momentum tensor
between glueball states and the vacuum
in SU(3) lattice gauge theory and extrapolate them
to the continuum.
These matrix elements may play an important phenomenological role
in identifying glue-rich mesons. 
Based on a relation derived long ago
by the ITEP group for $J/\psi$ radiative decays,
the scalar matrix element leads to a 
branching ratio for the glueball that is at least three times larger
than the experimentally observed branching ratio for the $f_0$
mesons above 1GeV.
This suggests that the glueball component must be diluted 
quite strongly among the known scalar mesons.
Finally we review the current best continuum determination
of the scalar and tensor glueball masses, the deconfining 
temperature, the string tension and the Lambda parameter, 
all in units of the Sommer reference scale,
using calculations based on the Wilson action. 
}

\begin{document}
\section{Introduction}
SU(N) gauge theories in 3+1 dimensions 
have been studied by lattice Monte-Carlo
techniques a long time~\cite{Creutz:1979dw}. 
Many low-energy dimensionless quantities
are now known at the percent level for $N=3$, as we shall review.
However, low-energy matrix elements of local operators
have not been given much attention. This is mainly because
the lowest possible dimension for a local gauge invariant operator
is four. For Monte-Carlo simulations, 
asymptotic freedom then implies that the amount of statistics
has to grow like $a^{-4}$ to maintain fixed
relative errors on the matrix elements~\cite{Meyer:2007ed} 
($a$ is the lattice spacing). This comes on top of the 
unavoidable $a^{-4}$ cost of simulating a four-volume fixed 
in physical units. In particular it is very expensive 
to take the continuum limit of renormalized matrix elements
of the energy-momentum tensor, 
as is well known to thermodynamics practitioners.
In this paper we compute the matrix elements of the energy-momentum 
tensor between the vacuum on the left and a
scalar or tensor glueball on the right. By using 
a \emph{locally} (i.e. on the scale 
of one lattice spacing)  smoothened gauge field,
we are able to reduce the prefactor of the $\sim a^{-4}$
cost function to a manageable size.

While these matrix elements have been computed 
previously~\cite{Chen:2005mg}, our technology
differs. In particular our lattice spacings 
are significantly  smaller and we improve on the statistical accuracy
of the matrix elements in the continuum by a factor of about two.
This improvement is due mainly to the reduced uncertainty on
the  non-perturbative normalization factors of the 
energy-momentum tensor.

One of our motivations is the possibility to confront
model predictions with the lattice data. Indeed
there are QCD sum rule predictions 
for these matrix elements~\cite{Novikov:1979va}.
Furthermore, models of QCD based on the 
AdS/CFT correspondence readily predict
many stables glueballs~\cite{Brower:2000rp},
 but the spectrum of the SU(3) theory
contains only two scalar and two tensor strictly stable 
glueballs, and this is  likely true 
at any finite number of colors $N$. By contrast, the detailed 
properties of the low-lying states, beyond their mass, 
represent opportunities for unambiguous comparisons.
The ability of the energy-momentum tensor to annihilate 
a glueball is a quantity of this type.  Such matrix elements
are somewhat analogous to $F_\pi$, which determines 
the width for the $\pi^+\to \mu^+\nu$ decay. 

Two other applications will be discussed in section 4. 
A long time ago the ITEP group derived 
an approximate expression~\cite{Novikov:1979va} for the 
branching ratio for a $J/\psi$ to decay into a scalar glueball
in terms of the matrix element that we compute in this paper.
We can thus roughly estimate the expected production rate
and compare to the experimental branching ratios for scalar mesons.

Secondly we show how the tensor and scalar matrix elements
can be used to constrain the thermal spectral functions
that determine respectively the shear and bulk viscosity of 
the plasma of gluons~\cite{Meyer:2007ic,Meyer:2007dy}.

In section 2 we define the matrix elements to be computed, 
and give the relation between the lattice observables
and the continuum, relativistically covariant quantities.
Section 3 describes the lattice calculation. In section 4 
we compare our results with those of~\cite{Chen:2005mg} and 
present the aforementioned applications. We end with 
a summary of the current knowledge of the low-energy 
properties of SU(3) gauge theory.

\section{Definitions}
In this section we fix our notation, 
define the relevant matrix elements and show how they
can be obtained from Euclidean correlation functions.

\subsection{Glueball matrix elements in the continuum}
Decomposing the energy-momentum tensor $T_{\mu\nu} $
into a traceless part $\theta_{\mu\nu}$ and a scalar part $\theta$ via
$T_{\mu\nu} = \theta_{\mu\nu} + \quarter \delta_{\mu\nu} \theta $,
the explicit Euclidean expressions are
\be
\theta(x) \equiv  \beta(g)/(2g) ~ F_{\rho\sigma}^a(x)  F_{\rho\sigma}^a(x)
\qquad\qquad
\theta_{\mu\nu}(x) \equiv
{\txts\frac{1}{4}}\delta_{\mu\nu}F_{\rho\sigma}^a F_{\rho\sigma}^a
   - F_{\mu\alpha}^a F_{\nu\alpha}^a .
\ee
The beta-function is defined by $qd\bar g/dq=\beta(\bar g)=-\bar g^3(b_0+b_1\bar g^2+\dots)$
and $b_0=11N/(3(4\pi)^2)$, $b_1=34N^2/(3(4\pi)^4 )$ in the SU($N$) pure gauge theory.
The gauge action reads $S_{\rm g}=\quarter F_{\mu\nu}^a F_{\mu\nu}^a$ in this notation.

We take over the notation of~\cite{Chen:2005mg} and define the matrix elements
\ba
\<\,\Omega | \theta(x) | S,p\,\>_{\rm r} 
&=& s ~e^{-ip\cdot x} 
\la{eq:s}\\
\<\,\Omega | \theta_{12}(x) | T,p,\sigma_-\,\>_{\rm r}&=& 
\<\,\Omega | \half(\theta_{11}-\theta_{22})(x) | T,p,\sigma_+\,\>_{\rm r}
= t ~e^{-ip\cdot x},~~~ {\bf p}=(0,0,p_3),
\la{eq:t}
\ea
where $|S\>$ and $|T\>$ respectively refer to the lightest scalar and tensor glueball states.
The labels $\sigma_\pm$ refer to the superpositions of helicity states
$|\sigma_+\>\propto |+2\>+|-2\>$ and $|\sigma_-\>\propto |+2\>-|-2\>$.
We define $s'$ and $t'$ in the same way for the first excited glueball in each of these
channels. The subscript `r' (relativistic)
indicates that the state has a Lorentz-invariant normalization,
\be
_{\rm r}\<\,U\, p\,\sigma|\,U\,q\,\sigma'\>_{\rm r} = 
2p^0 (2\pi)^3 \delta({\bf p-q})\,\delta_{\sigma\sigma'},
~~~~~ U=S,T.
\ee

\subsection{Glueball matrix elements from Euclidean correlation functions}
We now discuss  the extraction of glueball matrix elements  from 
correlation functions in the Euclidean theory,
 set up in a finite (but large) spatial volume.
Separately for the scalar and the tensor channels, 
we consider the correlation matrix 
\be
A_{ij}(\tau) = L^{-3}~ \<\, {\cal O}_i(0)~{\cal O}_j(\tau)\,\>_c,\qquad
i,j=0,1,\dots N_o.
\la{eq:cormat}
\ee
The subscript `c' indicates that we are dealing with the connected part
and $\tau$ is the Euclidean time variable.
Let us consider first the scalar channel.
The operator with label 0 is the definite-momentum projected local current,
\be
{\cal O}_0(\tau) = \int d^3x  ~ e^{i{\bf p\cdot x}}\,\theta(\tau,{\bf x}).
\ee
Operators 1 through $N_o$ are definite-momentum glueball operators, 
typically extended and 
designed to have large overlaps on the lightest two states.
In finite volume the spectral representation of (\ref{eq:cormat}) reads
\be
A_{ij}(\tau) = 
\sum_{n=1}^\infty ~\<\Omega|{\cal O}_i|n\> \, \<n|{\cal O}_j|\Omega\> ~e^{-E_n\tau}
\ee
where states are normalized as in quantum mechanics, $\<n|m\> = \delta_{nm}$
and have momentum ${\bf p}$ by momentum conservation.
In the infinite volume limit,  the connection between
the first state $|1\>$ 
with the one-particle state $|S,p\>_{\rm r}$ introduced above is
\be
\sqrt{2E_{\bf p} L^3}~ |1\> ~\to ~ |S,p\>_{\rm r}\,,
\ee
where $E_{\bf p}^2= M_S^2+{\bf p}^2$.
In particular, at large Euclidean time $\tau$, 
\be
A_{00}(\tau) = F_S^2 \,e^{-E_{\bf p}\tau} 
\,+\, F_{S'}^2\,e^{-E_{\bf p}'\tau} 
\,+\, {\rm O}(e^{-E_{\bf p}''\tau}).
\ee
with 
\be
F_S(L) \equiv L^{-3/2} |\<0|{\cal O}_0|1\>| 
\qquad
{\rm and}
\qquad
\lim_{L\to\infty} F_S(L)  = \frac{s}{\sqrt{2E_{\bf p}}}
\ee
and similarly for the excited state $|2\>$ which in the infinite volume
limit has energy $E_{\bf p}'^2= M_{S'}^2+{\bf p}^2$.

In the tensor channel, similar  equations apply as long as
${\bf p}$ is collinear with the polarization axis. 
On the lattice, the equality between the two forms of \eq\ref{eq:t}
is violated by O($a^2$) discretization errors as well as finite-size effects.
Here we use the second form exclusively, but we perform checks
for both sources of systematic error.

In the following we will only use ${\bf p}=0$.

\section{Lattice calculation} 
We simulate the SU(3) gauge theory using the 
Wilson plaquette action~\cite{Wilson:1974sk} at three values 
of the bare coupling, $\beta=6/g_0^2=6.0$, 6.2 and 6.408.
This corresponds to values of the Sommer parameter 
$r_0/a=5.368(35)$, $7.383(55)$ and $9.845(85)$~\cite{Necco:2001xg}.
The lattice size is respectively $16^3\times24$, $20^4$ and $28^4$.
On the coarsest lattice, we also check for finite-volume
effects by performing an additional simulation
on a $20^3\times24$ lattice.
The glueball spectrum was determined rather accurately
in~\cite{Meyer:2004gx} at these lattice spacings.
At the smallest lattice spacing, we apply a conversion
factor (given by the ratio of $r_0$ values)
to convert the spectrum from $\beta=6.4$ to $\beta=6.408$.
We use the standard combination of heatbath and 
overrelaxation~\cite{Creutz:1980zw,Cabibbo:1982zn,Kennedy:1985nu,Fabricius:1984wp} 
sweeps for the update in a ratio increasing from 
3 to 5  as the lattice spacing is decreased.
The overall number of sweeps between measurements was also
increased,  from 8 to 12.

We use the `HYP-plaquette' discretization of the energy-momentum
tensor, as described in~\cite{Meyer:2007tm}. 
We determine its normalization 
non-perturbatively, following the same strategy as in~\cite{Meyer:2007tm}.
The normalization of the discretization based on the bare plaquette is 
fixed by lattice sum rules, as is well-known in the context of 
thermodynamics~\cite{Engels:1980ty}. It is therefore sufficient to 
determine the normalization of the discretization based on the 
HYP-plaquette, which we employ here, relative to the bare plaquette.
A straightforward way to do this is to match the pressure and energy 
density computed with either discretization at a specific temperature.
The choice $T=1.21T_c$ made in ~\cite{Meyer:2007tm, Meyer:2007ed} insures
that a large signal is obtained for
$\<\theta\>$ and that  $T^{-4}\<\theta_{00}\>$ is 
already more than half its Stefan-Boltzmann limit.
Secondly, this choice implies $\Nt\geq6$ for $\beta\geq6.0$, so 
that large cutoff effects are avoided.
These relative normalization factors, denoted by $\chi_s(g_0)$ and $\chi(g_0)$
respectively for $\theta$ and $\theta_{00}$, are given in table~\ref{tab:chi}.
\TABLE[t]{
\begin{tabular}{c@{~~}c@{~~}c}
\toprule
$\beta$    &  $\chi_s(g_0)$      &  $\chi(g_0)$     \\
\midrule
6.000      &  0.9951(77)  &  0.5489(68)   \\
6.093      &  0.9778(89)  &  0.546(14)    \\
6.180      &  0.976(12)   &    0.596(20)   \\
6.295      &  0.953(20)   &    0.563(28)  \\
6.408      &  0.985(42)   &   0.612(49)     \\
\bottomrule
\end{tabular}
~~
\begin{tabular}{c@{~~}c@{~~}c@{~~}c@{~~}c}
\toprule
$\beta$    &  $aM_S$      & $aM_S^*$     &  $aM_T$      & $aM_T^*$    \\
\midrule
6.000      &  0.7005(47) & 1.167(25)   & 1.0596(64)  & 1.433(14)   \\
6.100      &  0.6021(85) & 1.038(15)   & 0.916(11)   & 1.180(34)   \\
6.200      &  0.5197(51) & 0.929(10)   & 0.7784(79)  & 1.032(20)  \\
6.400      &  0.3960(93) & 0.690(18)   & 0.5758(32)  & 0.795(28)  \\
\bottomrule
\end{tabular}
\caption{Left: normalization factors for the 
HYP-smeared plaquettes relative to the bare plaquette.
Right: the glueball masses from~\cite{Meyer:2004gx} relevant to this work.}
\la{tab:chi}
}
The data is conveniently parametrized as
($6 \leq 6/g_0^2 \leq 6.408$)
\ba
\chi_s(g_0) &=& 0.9731 + 0.67(g_0^2-6/6.18) \\
\chi(g_0)  &=&  0.5701-0.77(g_0^2-6/6.18).
\ea
In both cases, the absolute error increases from 0.007 at $\beta=6$ to
0.020 at $\beta=6.408$.
We extract $dg_0^{-2}/d\log a$ from the parametrization of $\log(r_0/a)$
given in~\cite{Necco:2001xg} and use the parametrization of $Z(g_0)$
given in~\cite{Meyer:2007ic}.

We employ linear combinations of Wilson loops that project in the 
$A_1^{++}$ and $E^{++}$ irreducible representations of the cubic group.
They are constructed from spatial links variables, 
using smearing and blocking as described in~\cite{Lucini:2004my}.

\subsection{Extraction of the glueball matrix elements}
The glueball matrix elements
can be extracted at sufficiently large $\tau$ using the fit ansatz
\ba
\widehat A_{ij}(\tau) &=& \sum_{n=1,2} ~ c^{(i)}_n \, c^{(j)}_n e^{-M_n\tau}
\la{eq:fit}
\\
\Rightarrow \quad
F_{U} &=&  |c^{(0)}_{1}|\,,  \qquad  F_{U'} =  |c^{(0)}_{2}|,
\qquad U=S,T.
\ea
A few remarks are in order:
\bi
\item the correlator $A_{00}$ is not included in the fit.
\item in our discretization the operator ${\cal O}_0$ is defined 
at half-integer times (in lattice units); thus
for $i\geq1,j=0$, $\tau/a$ takes half integer values in (\ref{eq:fit}),
whereas for $i,j\geq1$ it takes integer values.
\item
the glueball spectrum is already accurately known at the 
same simulations parameters from (\cite{Meyer:2004gx}, Table 7.1). 
We reproduce the values $\overline M_n$ in table~\ref{tab:chi}
and treat them as a `prior'.
\ei
Note however that since this 
information is exact within its quoted error, it is unnecessary 
to invoke Bayesian arguments to make use of it.
Therefore we determine the fit parameters by minimizing
 $\chi^2=\chi^2_{\rm I}+\chi^2_{\rm II}$, with
\ba
\chi^2_{\rm I} &=& \sum_{\stackrel{i\leq j}{ k\leq l}}\,
\sum_{\tau=u_{ij}}^{v_{ij}}\, \sum_{\tau'=u_{kl}}^{v_{kl}} 
 [A_{ij}(\tau)- \widehat A_{ij}(\tau)] 
~ {C^{-1}}_{(ij\tau),(kl\tau')} 
~[A_{kl}(\tau')- \widehat A_{kl}(\tau')]
\la{eq:chi1} \\
\chi^2_{\rm II} &=& \sum_{n=1,2} \frac{(M_n-\overline M_n)^2}{\sigma_n^2}\,.
\la{eq:chi2}
\ea

We use the full correlation matrix $C$ in this fit. Autocorrelation
effects in Monte-Carlo time were found to be negligible, 
so that $C_{(ij\tau),(kl\tau')}$  could be taken 
to be the usual estimator of 
$N_{\rm mst}^{-1}[\<A_{ij}(\tau)A_{kl}(\tau')\>-\<A_{ij}(\tau)\>\<A_{kl}(\tau')\>]$,
where $\<.\>$ are path integral averages and $N_{\rm mst}$ is the number of measurements.
We estimated the relative error on the eigenvalues of $C$ and found them to be 
small, due to the high statistics of the calculation. 
This makes the inversion of $C$ a stable procedure.

\FIGURE[t]{
\psfig{file=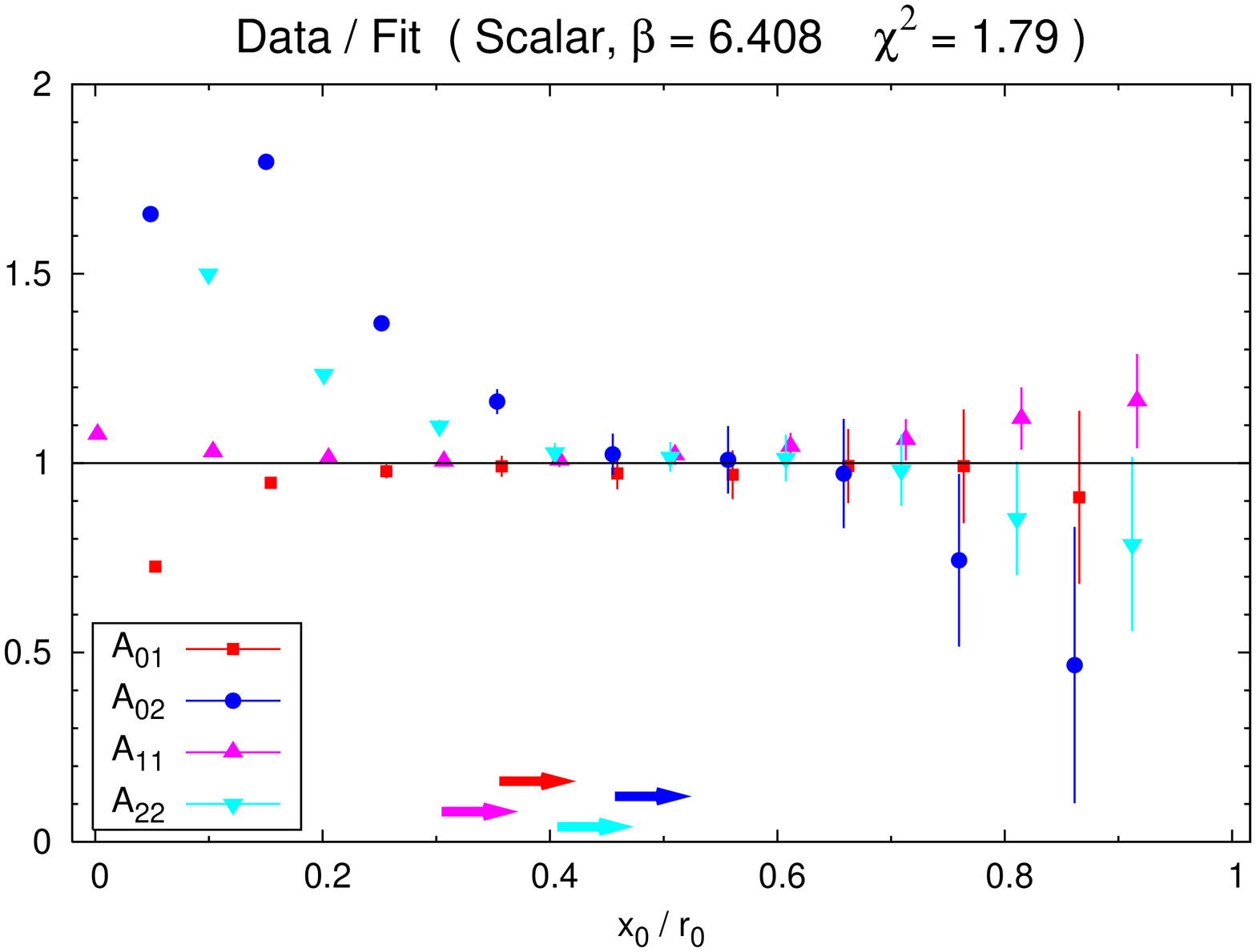,angle=-90,width=13cm}
\psfig{file=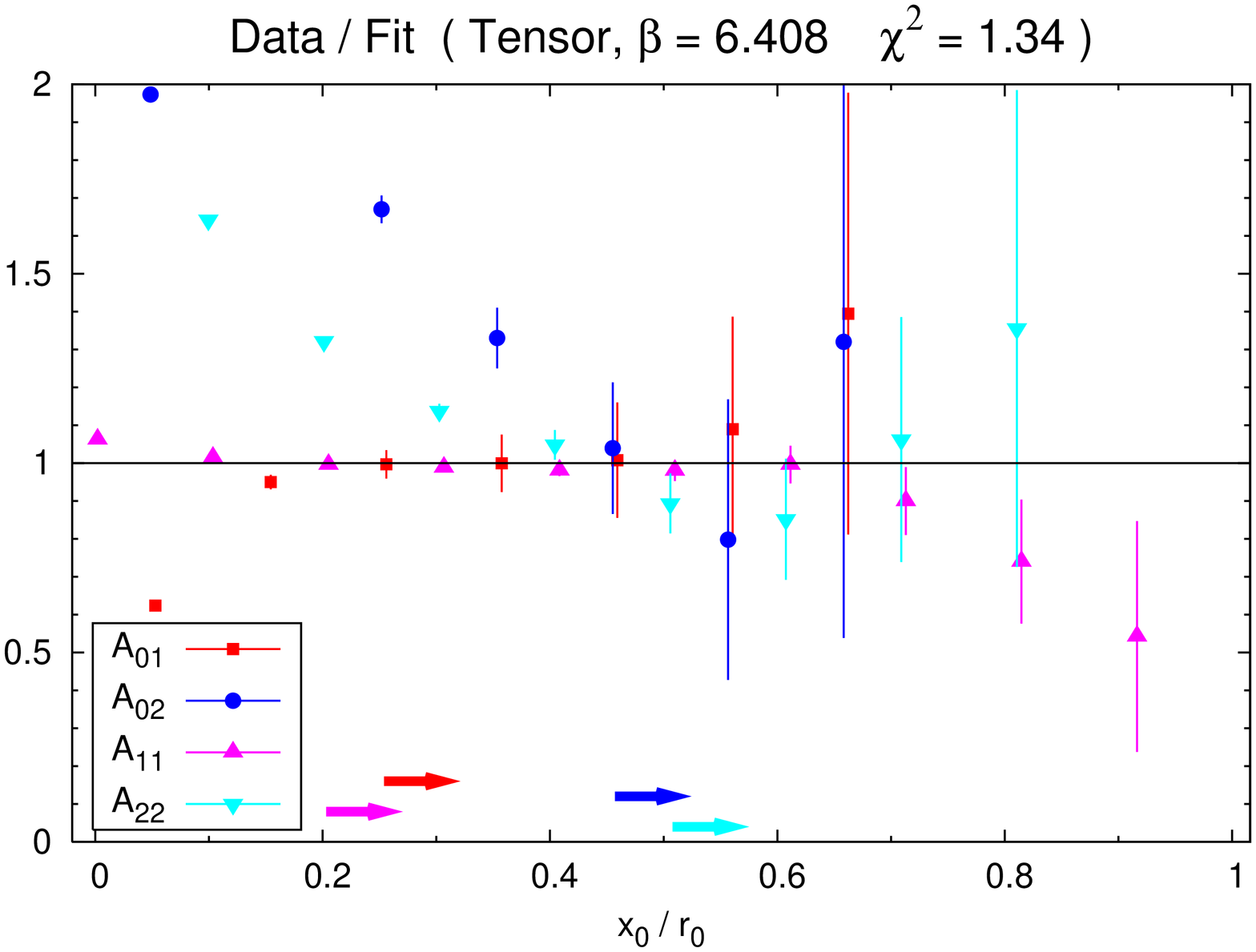,angle=-90,width=13cm}
\caption{The ratio of the scalar (top) and tensor (bottom) 
correlators to the fit, 
\eq\ref{eq:fit}. The arrows indicate for each correlator
where the fit starts.}
\la{fig:408}
}

We used two non-local operators in the fits, i.e. the indices $i,j,k,l\leq 2$ 
in \eq\ref{eq:chi1}. These operators were  linear combinations of non-local
operators in the relevant lattice irreducible representation 
designed to have a good projection on the lightest two states.
Increasing the number of operators in the fit did not seem 
to improve the determination of the physical parameters.
The results and minimized $\chi^2$ of these fits are given in Table~\ref{tab:F}.
The $\chi^2$ are of order unity, with a tendency to be slightly larger that one.
This may be due (partly) to the fact that the standard estimator for the $\chi^2$ 
we are using is an upward-biased one~\cite{ulli-notes}. Beyond the value of the 
$\chi^2$ it is important  to check that the fit is stable against variations 
in the fit ranges $u_{ij}\leq \tau \leq v_{ij}$, and that no strong trend 
is seen in $A_{ij}(\tau)/\widehat A_{ij}(\tau)$ as a function of $\tau$ within 
the fit range. 
Figure~\ref{fig:408} displays this ratio in the scalar and in the tensor sectors
for the simulation at the smallest lattice spacing.
We see that the correlators $A_{01}$ and $A_{11}$ have small deviations
from the fit even outside the fit range. This gives us confidence that
excited state ($n\geq3$) contaminations are negligible in the fit range.
This property is much less satisfied for the $A_{02}$ and $A_{22}$ correlators.
Although the fit range for the correlators starts further out in $\tau$, 
this implies that the control over excited state contributions is much
less good for $F_{S^*}$ and $F_{T^*}$. We will therefore not include 
these matrix elements in our final list of results.

\FIGURE[t]{
\psfig{file=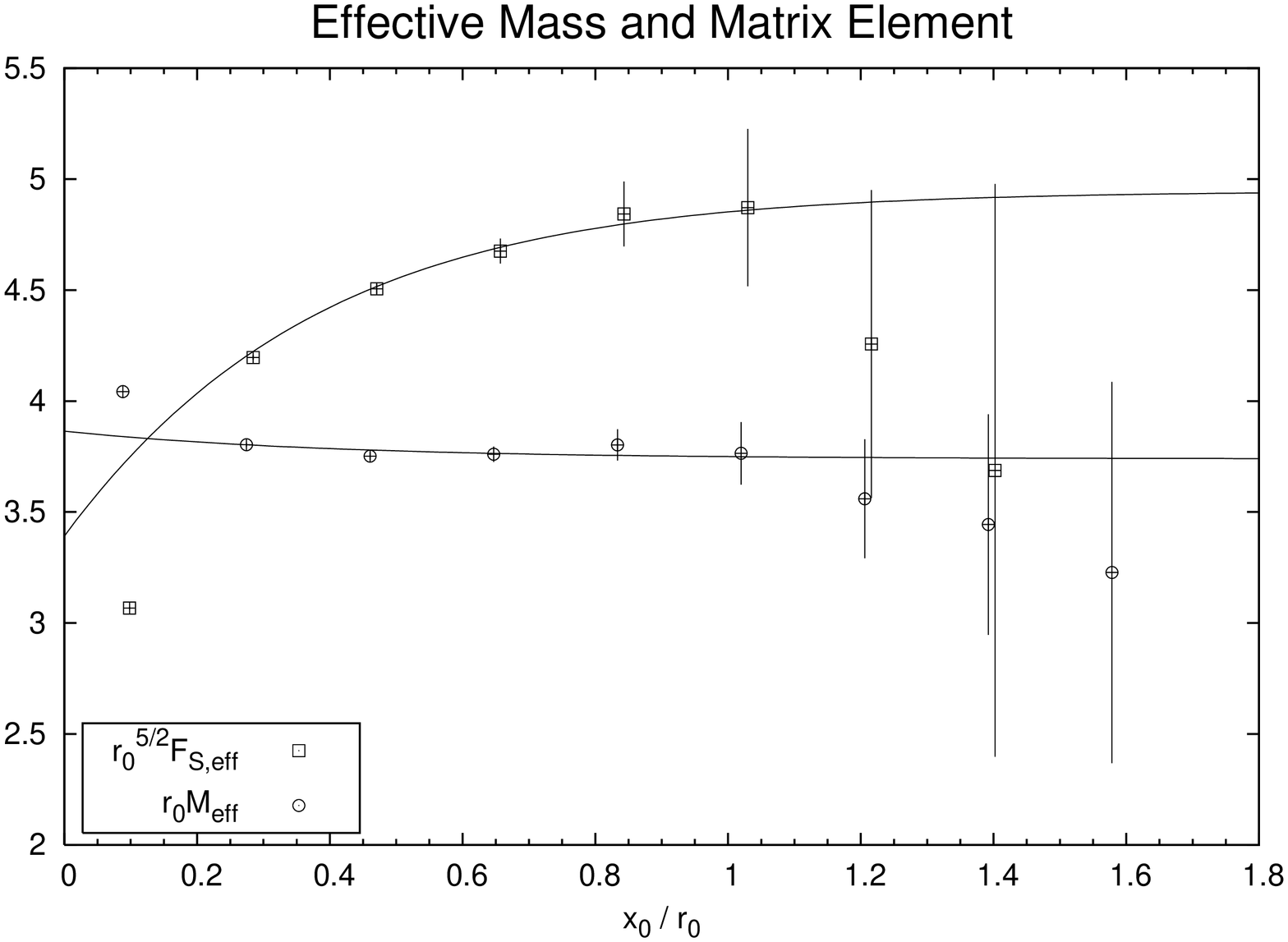,angle=-90,width=13cm}
\caption{The effective mass and matrix element in the scalar sector 
at $\beta=6$, $16^3\times24$ lattice. The fit \eq\ref{eq:fit} is shown.
The fit starts at $x_0=a\approx0.19r_0$ for the diagonal correlator and at 
$x_0=\frac{5}{2}a\approx0.47r_0$ for the correlator 
between the local and the smeared operator.}
\la{fig:eff}
}

We may use the effective mass and the effective matrix element
\ba
am_{\rm eff}(t)&=&\log \frac{A_{11}(\tau)}{A_{11}(\tau+a)},
\la{eq:Meff}
\\
F_{U,{\rm eff}}(\tau-\half a)  &=& A_{01}(\tau-\half a) 
\frac{A_{11}(\tau-a)^{(\tau-a)/2a}}{A_{11}(\tau)^{\tau/2a}},
\qquad U=S,T,\qquad \tau=2a,3a,\dots
\la{eq:Feff}
\ea
as a  way to visualize the contributions of excited states to the 
physical quantities we extract (see \fig\ref{fig:eff}).
The effective quantities computed from 
the fit to the correlators (\eq\ref{eq:chi1},\ref{eq:chi2}) are also shown 
as curves on the plot. The fit looks convincing. 
We find that, in general, the method of determining the 
mass and matrix elements from the effective quantities is 
less stable than fitting the correlators. 
A possible reason for this is that
the histograms of ratios as in \eq\ref{eq:Feff} can become
arbitrarily non-Gaussian if the denominator is noisy
(e.g. at large $\tau$), even if
the correlators were perfectly Gaussian distributed.

Finally, increasing the volume from $16^3$ to $20^3$ at $\beta=6.0$ 
does not affect the matrix elements in a statistically significant way (see
Tab.\ref{tab:F}). An effect at the subpercent level is seen on 
some of the glueball masses, because they are so accurately determined,
but this does not affect the discussion in the rest of this paper.

\TABLE[t]{
\centerline{
\begin{tabular}{c@{~~}c@{~~}c@{~~}c@{~~}c}
\toprule
$\beta$    &   $aM_{\rm S} $    &  $aM_{\rm S^*}$ 
               &  $aM_{\rm T}$ & $aM_{\rm T^*}$  \\
\midrule
6.000 ($20^3$) & 0.7032(16) & 1.1733(37)  & 1.0587(15) & 1.43261(53) \\
6.000 ($16^3$) & 0.6966(16) & 1.1747(58)  & 1.0592(22) & 1.4358(10)  \\
6.200          &0.51918(96) & 0.93010(30) & 0.7783(20) & 1.0399(22)  \\
6.408          &0.39603(63) & 0.6946(11)  & 0.5937(36) & 0.8322(92)  \\
\bottomrule
\end{tabular}}

\\
\\
\\

\begin{tabular}{c@{~~}c@{~~}c@{~~}c@{~~}c@{~~}c@{~~}c}
\toprule
$\beta$    &   $F_{\rm S}r_0^{5/2} $    &  $F_{\rm S^*}r_0^{5/2}$ & $\chi^2/\nu$
               &  $F_{\rm T}r_0^{5/2}$ & $F_{\rm T^*}r_0^{5/2}$  & $\chi^2/\nu$  \\
\midrule
6.000 ($20^3$) & 5.05(06)(09) & 6.53(30)(12) &1.33  & 3.63(26)(08)  &  8.49(78)(27) & 1.29\\
6.000 ($16^3$) & 4.95(03)(09) & 6.21(19)(12) &1.36  & 3.41(11)(08)&  7.72(40)(25) & 0.94 \\
6.200          & 4.43(25)(09) & 10.25(90)(22)& 1.50 & 3.07(32)(09)& 10.37(68)(29) & 1.29\\
6.408          & 4.55(36)(14) & 9.75(54)(31) &1.79  & 2.30(33)(10)& 11.04(38)(48)& 1.34 \\
\bottomrule
\end{tabular}
\caption{The scalar (S) and tensor (T) 
glueball masses and matrix elements extracted from the fits. In the lower
table, the first error 
is the uncertainty coming from the bare matrix element in lattice units, 
the second is the cumulated error of all the other factors entering 
the renormalization group invariant quantity.}
\la{tab:F}
}

\subsection{Continuum extrapolation}

\FIGURE{
\psfig{file=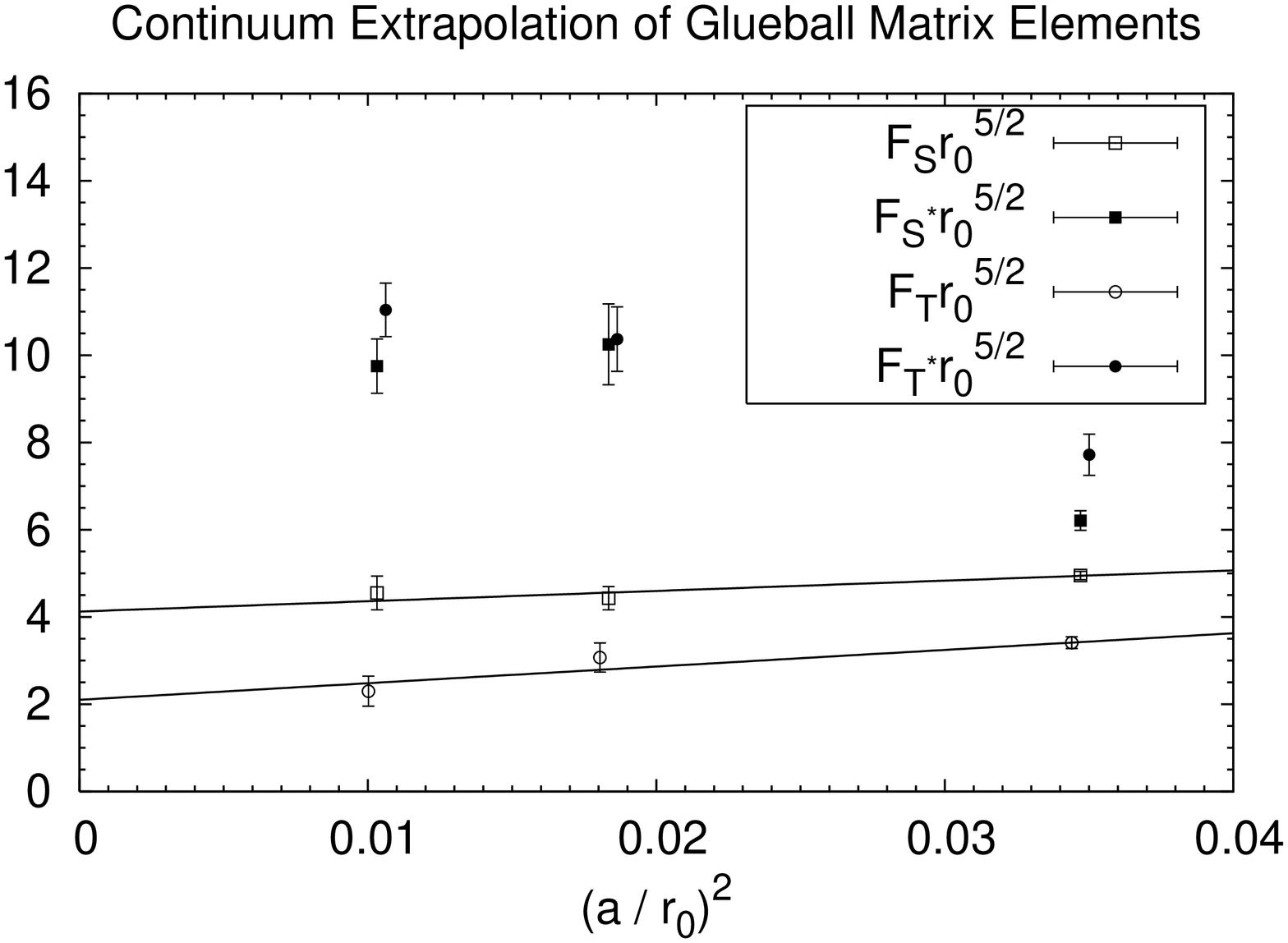,angle=-90,width=13cm}
\caption{Continuum extrapolation.}
\la{fig:conti}
}

Having obtained the matrix elements at three different lattice 
spacings, we can attempt a continuum extrapolation linear 
in $(a/r_0)^2$, as illustrated on \fig\ref{fig:conti}.
We find 
\ba
F_{\rm S}r_0^{5/2} &=&  4.13(40)\,,\qquad \chi^2/{\rm d.o.f.}=0.46
\\
F_{\rm T}r_0^{5/2} &=&  2.10(41)\,,\qquad \chi^2/{\rm d.o.f.}=0.99
\ea
The extrapolations are satisfactory: the $\chi^2$ is small, 
the slopes of the extrapolations are small to moderate. 
In particular the continuum value is statistically compatible
with the value from the smallest lattice spacing.

We do not attempt an extrapolation for the first excited states.
Indeed it is clear from  \fig\ref{fig:conti} that 
the $\beta=6.0$ data does not lie in the $a^2$ scaling region.
The values from the lighter two lattice spacings are consistent
with eachother, suggesting that cutoff effects are small beyond
$\beta=6.2$. Our best estimate of $F_{\rm S'}r_0^{5/2}$ 
and $F_{\rm T'}r_0^{5/2}$ are thus the values at the smallest 
lattice spacing, but one should keep in mind that our control
over excited state contamination is far less good for these
matrix elements; the induced systematic uncertainty is 
comparable or larger than the statistical error.

\section{Applications and summary}

In the following, we compare our results to previously
obtained ones. Then we present two applications 
of the computed matrix elements. We end with a summary
of our knowledge of non-perturbative dimensionless ratios 
in the SU(3) gauge theory.

\subsection{Technical comparison}
We can compare our results to those of~\cite{Chen:2005mg}.
We find\footnote{As compared to~\cite{Chen:2005mg}, 
our $s$ contains  an extra factor of $11/(4\pi)^2$  (coming 
from the beta-function).},
\ba
  && ~\textrm{this work} ~~~~~~ \textrm{Chen et al.} \\
sr_0^3 &=&     11.6(1.1)     \quad\quad  15.8(3.2)    
\\
tr_0^3 &=&    ~ 7.1(1.4)       \quad\quad  7.5(2.8)   
\ea
Comparing the matrix elements in units of $r_0$ saves us from
having to settle on a value for $r_0$ in fm 
($r_0^{-1}=410$MeV was used in \cite{Chen:2005mg}). 
We conclude that our results are in satisfactory agreement with 
those of \cite{Chen:2005mg}, and the statistical uncertainties have 
been reduced by a factor of at least two. 

We stress that while our continuum extrapolation 
uses data at lattice spacings $(a/r_0)^2$ in the range 0.010 to 0.035, 
the continuum extrapolation of~\cite{Chen:2005mg}
uses data in the range of spatial lattice spacings 0.044 to 0.22. 
The control of the continuum limit is thus qualitatively different.
One might worry that the support of the field strength operators
used for the `type II' discretization in~\cite{Chen:2005mg} varies
between 0.4fm and 0.9fm. On the other hand, the authors do find
good agreement between certain observables which only become 
strictly equal in the continuum limit.

We have used non-perturbative renormalization factors at 
every lattice spacing, while the procedure used in~\cite{Chen:2005mg}
for $s$ and $t$ implies that the continuum is approached asymptotically with
O($g_0^2$) as well as O($a^2$) corrections. 

Simulating at small lattice spacings comes at a heavy computational price
for these observables: the signal-to-noise ratio for matrix elements
of the gluonic dimension four operators decreases
with the fourth power of the lattice spacing~\cite{Meyer:2007ed}. 
This explains why 
our statistical errors on the bare matrix elements are somewhat larger
than in~\cite{Chen:2005mg}. 

The use of a coarse spatial lattice allowed the authors of~\cite{Chen:2005mg} 
to reach volumes significantly  larger than ours in physical units. 
We have however checked for finite volume effects explicitly, 
and find no significant variation in the matrix elements. 
It is known~\cite{Meyer:2004vr} that the 
low-lying spectrum of glueballs exhibits remarkably small finite-size effects
for $L>2.5r_0$ (and for certain channels they remain small for even smaller
box sizes).

\subsection{Glueball production rate in radiative $J/\psi$ decays}
First of all, we can compare our result for $s$ with 
the QCD sum rule prediction~\cite{Novikov:1979va} 
\be
s \approx \frac{11}{4\pi}\,\sqrt{\frac{G_0}{2b_0}}\,  M_S\,.
\ee
Using $G_0=0.012{\rm GeV}^4$ (the `gluon condensate')
and $r_0^{-1}=410$MeV this leads to
 $sr_0^3\approx6.0$, almost a factor two smaller than
our result. Since we regard the gluon condensate as 
a phenomenological parameter, this disagreement 
 does not surprise us too much.

We may use our value  for $s$ to 
estimate the partial width of $J/\psi$ to radiatively decay into 
a scalar glueball. An approximate expression for this 
experimentally observable quantity was derived in~\cite{Novikov:1979va},
\be
\Gamma(J/\psi\to G_0\gamma)\approx 
\frac{8\pi\,\alpha^3}{5^2\cdot11^2\cdot3^8}\,
\frac{m_{J/\psi}^4~s^2}{m_c^8\,\Gamma(J/\psi\to e^+e^-)}.
\ee
Here $m_c\simeq1250$MeV is the charm  mass in the $\overline{\rm MS}$ scheme 
at scale $\mu=m_c$, 
$m_{J/\psi}\simeq3097$MeV is the $J/\psi$ mass, $\alpha\simeq\frac{1}{137}$ 
is the fine structure constant
and $\Gamma(J/\psi\to e^+e^-)\simeq5.55$keV.
The ingredients that go into this formula are the $1/m_c$
expansion, a dispersion relation for the charm-quark loop induced
$\gamma\gamma\to gg$ transition and the assumption that 
the $J/\psi$ contribution dominates the spectral integral.
This leads to the estimate
\be
{\rm Br}(J/\psi\to G_0\gamma)\approx 0.009.
\la{eq:JtoG}
\ee

The authors of~\cite{Novikov:1979va} also suggest 
that a more accurate prediction is 
\be
{\rm Br}(J/\psi\to G_0\gamma) = 
\left(\frac{3\pi}{11} \right)^2 \,
\frac{s^2\cdot {\rm Br}(J/\psi\to\eta'\gamma)}
{|\<\Omega|\alpha_s F_{\mu\nu}^a\tilde F_{\mu\nu}^a|\eta'\>|^2}.
\ee
Here the assumption about the dominance of the $J/\psi$ 
is replaced by the assumption that the relative size
of the $J/\psi$ and the `continuum' contributions are 
identical in the scalar and the pseudoscalar channels.
Using Eq. (5) of~\cite{Novikov:1979uy}, we can trade the properties of 
$\eta'$ for the corresponding ones for $\eta$ and use 
the $SU(3)_{\rm f}$ result~\cite{Novikov:1979uy}
\be
\<\Omega|\alpha_s F_{\mu\nu}^a\tilde F_{\mu\nu}^a|\eta\>
= \frac{4\pi}{3}\sqrt{\frac{3}{2}}\,f_\eta m_\eta^2
\ee
with $f_\eta\approx170$MeV. Then we obtain
\be
{\rm Br}(J/\psi\to G_0\gamma)\approx
\frac{3^3}{11^2\cdot2^3}\left(\frac{1-x'^2}{1-x^2}\right)^3
\frac{s^2\cdot {\rm Br}(J/\psi\to\eta\gamma)}{f_\eta^2 m_\eta^4}
=5.6(1.7)\cdot{\rm Br}(J/\psi\to\eta\gamma).
\la{eq:Br2}
\ee
Here $x=m_\eta/m_{J/\psi}$ and $x'=m_{\eta'}/m_{J/\psi}$.
Since ${\rm Br}(J/\psi\to\eta\gamma)=0.98(10)\cdot10^{-3}$~\cite{Amsler:2008zz},
we obtain a somewhat lower result than (\ref{eq:JtoG}).
This is still a rather large branching ratio. For instance, 
the PDG~\cite{Amsler:2008zz} gives 
${\rm Br}(J/\psi\to\gamma f_0(1710))\approx 1.5(3)\cdot10^{-3}$
if one adds up the contributions of the $\gamma K\overline K$, 
$\gamma\pi\pi$ and $\gamma\omega\omega$ channels. The production rate
of $f_0(1500)$ is even smaller. According to (\ref{eq:Br2}), 
if any of the 
$f_0(1370)$, $f_0(1500)$ or $f_0(1710)$ states had a 
glueball component close to unity, they would be produced more copiously than
observed. Therefore, \eq\ref{eq:Br2} suggests that 
the glueball component is quite strongly diluted among the three states.
This  conclusion is also reached by doing 
detailed parametrizations and fits to experimental data of the 
mixing pattern~\cite{Close:1996yc,Close:2005vf}.

\subsection{Constraining thermal spectral functions}
Another application arises in the calculation of 
transport coefficients in the plasma of gluons.
Indeed the local-current two-point function $A_{00}$ can be expressed 
in terms of the spectral function $\rho(\omega,{\bf p},T)$, 
\be
A_{00}(\tau,{\bf p},T)=
\int_0^\infty d\omega\,\rho(\omega,{\bf p},T) \,
\frac{\cosh \omega(\frac{1}{2T}-\tau )}{\sinh \frac{\omega}{2T}}\,.
\la{eq:rho}
\ee
Here $T$  is the  temperature. 
In the tensor channel the shear viscosity is then given by
\be
\eta(T) = \pi \lim_{\omega\to0} \frac{\rho(\omega,{\bf 0},T)}{\omega}\,.
\ee
Because $A_{00}$ is dominated 
by  ultraviolet contributions that are temperature independent,
in~\cite{Meyer:2008dq} we proposed to subtract from $A_{00}$ what that 
correlator would be if the spectral function was the same as at $T=0$,
before solving \eq\ref{eq:rho} for $\rho(\omega,{\bf p},T)$.
Let us call that would-be correlator $\widetilde A_{00}$.
The $T=0$ spectral function has a simple expression in terms of energy
levels and matrix elements of the kind we calculated in this paper,
\be
\rho(\omega,{\bf p},T=0)= L^{-3}\sum_n \delta(\omega-E_n)\,
 |\<0|{\cal O}_0 |n\>|^2.
\la{eq:rho0}
\ee
Therefore, we find 
that the contributions of the first two 
terms in \eq\ref{eq:rho0} to $\widetilde A_{00}(\tau,{\bf 0},T)$ 
read (at $N_t=1/(aT)=8$)
\ba
\widetilde A_{00}(\tau,{\bf 0},1.24T_c)
&=& 1.25 + 5.2 + \dots
\la{eq:1p24}
\\
\widetilde A_{00}(\tau,{\bf 0},1.65T_c)
&=& 0.388 +  3.7 +\dots
\la{eq:1p65}
\ea
These contributions are substantial since $A_{00}(\tau,{\bf 0},T)=8.05(32)$
and $8.73(33)$ respectively~\cite{Meyer:2008dq}. The first excited 
state contribution appears to be particularly large. Due to 
possible higher state contamination in our estimate of $F_{S^*}$ and $F_{T^*}$
(see section 3), 
it may be that the second terms in \eq\ref{eq:1p24} or \ref{eq:1p65}
effectively amounts to the contribution of more than one state. 

\subsection{A summary of the low-energy parameters of SU(3) gauge theory}
We finish with a summary of our knowledge of low-energy dimensionless 
quantities in the SU(3) gauge theory. The results we use were all obtained with the 
Wilson plaquette action~\cite{Wilson:1974sk} and we extrapolate them to the continuum.
We use the parametrization of $(r_0/a)(\beta)$ from~\cite{Necco:2001xg}.
If $\sigma$ is the string tension, $T_c$ the deconfining temperature
and $\Lambda_{\overline{\rm MS}}$ the Lambda parameter in the $\overline{\rm MS}$ scheme,
we find in the continuum
\ba
M_S\,r_0 &=& 3.958(47) \qquad \chi^2/(4-2)=0.09, \\
M_T\,r_0 &=& 5.878(77)   \qquad \chi^2/(4-2)=0.6,\\
s\,r_0^3   &=& 11.6(1.1) \qquad \chi^2/(3-2)=0.5,\\
t\,r_0^3   &=& 7.1(1.4) \qquad \chi^2/(3-2)=1.0,\\
\sqrt{\sigma}\,r_0 &=& 1.1611(95)  \qquad \chi^2/(5-2)=0.1,\\
T_c\,r_0 &=& 0.7463(64)  \qquad \chi^2/(4-2)=0.3, \\
\Lambda_{\overline{\rm MS}}\,r_0 &=& 0.60(5).
\ea
The data for  $aM_S$, $aM_T$ $sa^3$ and $ta^3$   is from this work,
plus the $\beta=6.1$ glueball mass data from~\cite{Meyer:2004gx}.
The string tension $a\sqrt{\sigma}$ is taken from~\cite{Lucini:2004my}
($\beta\geq 5.8000$) and from~\cite{Meyer:2004gx} ($\beta=6.4$).
We took the critical values of $\beta$ for given $N_t=1/(aT)$ 
from~\cite{Boyd:1996bx} ($N_t=6,8,12$) and from~\cite{Lucini:2003zr} ($N_t=5$).
The Lambda parameter value is the data of~\cite{Capitani:1998mq}.
We do a continuum extrapolation linear in $(a/r_0)^2$ for the first five quantities.
We extrapolate $T_cr_0$  as a function of $1/N_t^2$.
The quantity $\Lambda_{\overline{\rm MS}}r_0$ is already given in the 
continuum in~\cite{Capitani:1998mq}.

The overall level of accuracy achieved is remarkable.

\section{Conclusion}
In summary, we have computed scalar and tensor 
glueball matrix elements in SU(3) gauge theory with improved precision.
A straightforward application to the production of scalar glueballs
in $J/\psi$ radiative decays suggests that 
none of the known scalar mesons can contain too large 
a glueball component.
Finally we gave a summary of the current knowledge 
of seven non-perturbative observables of the SU(3) gauge theory.
We hope that these quantities constitute a useful set for models 
and semi-analytical methods to calibrate on and compare predictions to.

\acknowledgments{
The simulations were done on the desktop machines of the
Laboratory for Nuclear Science at M.I.T. 
This work was supported in part by
funds provided by the U.S. Department of Energy 
under cooperative research agreement DE-FG02-94ER40818.
}

\bibliographystyle{JHEP}
\bibliography{/afs/lns.mit.edu/user/meyerh/BIBLIO/viscobib}

\end{document}